\documentclass[12pt]{article}
\usepackage{epsfig,amsfonts,amssymb,amsbsy,amsbsy}
\def\cvp{\raise 2pt\hbox{,}}

\def\tr{\mathop{\rm tr}\nolimits}
\def\im{\mathop{\rm Im}\nolimits}

\def\plb#1#2#3{{\it Phys.\ Lett.\ }{\bf B #1} (#2) #3}
\def\npb#1#2#3{{\it Nucl.\ Phys.\ }{\bf B #1} (#2) #3}
\def\prl#1#2#3{{\it Phys.\ Rev.\ Lett.\ }{\bf #1} (#2) #3}
\def\jhep#1#2#3{{\it J. High Energy Phys.\ }{\bf #1} (#2) #3}

\def\cmp#1#2#3{{\it Comm.\ Math.\ Phys.\ }{\bf #1} (#2) #3}

\def\ap#1#2#3{{\it Ann.\ Phys.\ }{\bf #1} (#2) #3}
\def\u1{{\rm U}(1)}
\def\cpN{{\mathbb C}P^{N}}

\begin{document}
%
%
\pagestyle{empty}
{\parskip 0in
\hfill NEIP-01-010

\hfill PUPT-2007

\hfill LPTENS-01/36

\hfill hep-th/0111117}

\vfill
\begin{center}
{\LARGE A note on theta dependence}

\vspace{0.4in}

Frank F{\scshape errari}{\renewcommand{\thefootnote}{$\!\!\dagger$}
\footnote{On leave of absence from Centre 
National de la Recherche Scientifique, Laboratoire de Physique 
Th\'eorique de l'\'Ecole Normale Sup\'erieure, Paris, France.}}\\
\medskip
{\it Institut de Physique, Universit\'e de Neuch\^atel\\
rue A.-L.~Br\'eguet 1, CH-2000 Neuch\^atel, Switzerland\\
and\\
Joseph Henry Laboratories\\
Princeton University, Princeton, New Jersey 08544, USA}
\smallskip
{\tt frank.ferrari@unine.ch}
\end{center}
\vfill\noindent
The dependence on the topological $\theta$ angle term in quantum 
field theory is usually discussed in the context of instanton calculus. 
There the observables are $2\pi$ periodic, analytic functions of 
$\theta$. However, in strongly coupled theories, the semi-classical 
instanton approximation can break down due to infrared 
divergences. Instances are indeed known where analyticity in $\theta$ 
can be lost, while the $2\pi$ periodicity is preserved.
In this short note we exhibit a simple two dimensional example where 
the $2\pi$ periodicity is lost. The observables remain periodic under 
the transformation $\theta\mapsto\theta + 2 k\pi$ for some $k\geq 2$. We 
also briefly discuss the case of four dimensional ${\cal N}=2$ 
supersymmetric gauge theories.
\vfill
\begin{flushleft}
November 2001
\end{flushleft}
\newpage\pagestyle{plain}
\baselineskip 16pt
\setcounter{footnote}{0}
%
A topological $\theta$ angle term $L_{\theta}$ can be added to the 
lagrangian in 
various quantum field theories. The most important and well-known 
example is the case of four dimensional gauge theories, for which
\begin{equation}
\label{lt4d}
L_{\theta}^{4\rm D} ={\theta\over 32\pi^{2}}\,
\epsilon^{\mu\nu\rho\kappa}\tr F_{\mu\nu}F_{\rho\kappa}\, .
\end{equation}
Other examples include two dimensional gauge theories, for which
\begin{equation}
\label{lt2d}
L_{\theta}^{2\rm D} ={\theta\over 4\pi}\,\epsilon^{\mu\nu}F_{\mu\nu}\, ,
\end{equation}
or two dimensional non-linear $\sigma$ models with target space $\cal M$,
for which
\begin{equation}
\label{lts}
L_{\theta}^{\sigma}= {\theta\over 4\pi}\, \epsilon^{\mu\nu} 
B_{ij}(\phi)\partial_{\mu}\phi^{i}\partial_{\nu}\phi^{j}\, ,
\end{equation}
where $B\in H^{2}({\cal M},{\mathbb Z})$ has integer periods (for 
K\"ahler $\sigma$ models like the $\cpN$ model in which we will be 
interested in, $B$ is the suitably normalized K\"ahler form).
All the $\theta$ angle terms can be written locally
as total derivatives, and are normalized in such a way that with classical 
boundary conditions at infinity (determined by the requirement of finite 
classical action),
\begin{equation}
\label{qc}
\int\! L_{\theta}/\theta \in {\mathbb Z}\, .
\end{equation}
These two properties have two important consequences in {\it weakly 
coupled} field theories. The fact that $L_{\theta}$ is a total 
derivative implies that there is no $\theta$ dependence in 
perturbation theory. The fact that $\int\! L_{\theta}/\theta$ is 
quantized implies that topologically non-trivial classical field 
configurations, called instantons, that can contribute to the path 
integral in a semiclassical approximation, may induce a $2\pi$ 
periodic and smooth $\theta$ dependence. More precisely, a $k$ instanton or 
$k$ anti-instanton contribution, $k\in \mathbb Z$, is proportional to
\begin{equation}
\label{insc}
e^{-8\pi^{2}k/g^{2}}e^{\pm ik\theta}\, ,
\end{equation}
where $g$ is the conventionally normalized coupling constant. In 
theories where dimensional transmutation takes place, the running 
coupling $g$ is replaced by a scale $|\Lambda|$. It is convenient to 
introduce a complexified scale
\begin{equation}
\label{scale}
\Lambda = |\Lambda|e^{i\theta/\beta}\, ,
\end{equation}
where $\beta$ is the coefficient of the one-loop $\beta$ function. 
Contributions from $k$ instantons and $k$ anti-instantons in the one-loop 
approximation are then respectively proportional to 
$\Lambda^{k\beta}$ and $\bar\Lambda^{k\beta}$.

Understanding the $\theta$ dependence in {\it strongly coupled} field 
theories is a much more difficult problem. Boundary conditions at infinity, 
or equivalently the structure of the vacuum, should be determined by using 
the (generally unknown)
quantum effective action, not the classical action, and quantization 
laws such as (\ref{qc}) can be invalidated. Moreover, subtle infrared 
effects can induce a $\theta$ dependence in Feynman diagrams. The 
conclusion is that analyticity or $2\pi$ periodicity are not ensured
a priori. A typical example where the dependence in
$\theta$ is not consistent with instanton calculus
is the two dimensional quantum electrodynamics,
or equivalently the purely bosonic $\cpN$ non-linear
$\sigma$ model whose effective action is QED$_{2}$ \cite{adda}. There it is 
well-known that the term (\ref{lt2d}) induces a constant
background electric field $E=e^{2}\theta /2\pi$, where $e$ is the electric 
charge \cite{coleman}. The phenomenon of pair creation, which is 
energetically favoured for $|E|>e^{2}/2$, ensures $2\pi$ periodicity in 
$\theta$, but analyticity at $\theta=\pm\pi$ is lost. Discussions of 
similar effects in various contexts can be found for example in
\cite{added}.

An interesting generalization of the standard QED$_{2}$ discussed above
is the ${\cal N}=2$ supersymmetric QED$_{2}$ with a twisted 
superpotential $W$, or 
equivalently the ${\cal N}=2$ supersymmetric $\cpN$ model with 
twisted masses $m_{i}$ \cite{AG,HH,Dorey}. There, in 
addition to the gauge field $E=F^{01}$,
one has a Dirac fermion $\lambda$ and a 
complex scalar $\sigma$. All these 
fields belong to the same supersymmetry multiplet and are packed up in a 
single (twisted) chiral superfield
\begin{equation}
\label{ssuper}
\Sigma = \sigma -2i(\theta_{-}\bar\lambda_{+} + 
\bar\theta_{+}\lambda_{-}) + 2\bar\theta_{+}\theta_{-}(D-iE).
\end{equation}
The real auxiliary field $D$ appears on the same footing as the gauge 
field $E$, consistently with the fact that gauge fields do not propagate in 
two dimensions. The equation 
for the background electric field is in this case
$E=-2e^{2}\im W'(\sigma)$, which shows that in the supersymmetric vacua for 
which $W'(\sigma)=0$, there is actually no background electric field, 
whatever $\theta$ may be. The phenomenon of pair creation never occurs, 
since the supersymmetric vacua have zero vacuum energy and are thus stable.
What is left then of the $2\pi$ periodicity in $\theta$? For the $\cpN$ 
model, the twisted superpotential is \cite{adda2,HH,Dorey}
\begin{equation}
\label{weff2}
W(\sigma) = {\sigma\over 4\pi}\,\ln\prod_{i=1}^{N+1}{\sigma + 
m_{i}\over e\Lambda} + {1\over 4\pi}\sum_{i=1}^{N+1} m_{i}\ln 
{\sigma + m_{i}\over e\Lambda}\,\cvp
\end{equation}
where $\Lambda$ is given by (\ref{scale}) with $\beta=N+1$.
The field $\sigma$ is chosen so that $\sum_{i=1}^{N+1}m_{i}=0$. The 
vacuum equation $W'(\langle\sigma\rangle)=0$ reduces to
\begin{equation}
\label{vaceq}
\prod_{i=1}^{N+1} \bigl(\langle\sigma\rangle + m_{i}\bigr)
= \Lambda^{N+1}\, .
\end{equation}
This equation has $N+1$ solutions that correspond generically
to $N+1$ physically inequivalent vacua $|i\rangle$, $1\leq i\leq N+1$.
The question we want to address is then the following:
can we trust instanton calculus, and in particular the 
quantization condition (\ref{qc}), in those vacua?

An interesting 
property of the theory described by (\ref{weff2}), shared more 
generally by asymptotically free
non-linear $\sigma$ models with mass terms \cite{neu}, 
is that any of the vacua can be made arbitrarily weakly coupled by 
suitably choosing the mass parameters. If 
$|m_{i}-m_{k}|\gg |\Lambda|$ for all $i\not =k$, then the vacuum 
$|k\rangle$ characterized by $\langle k|\sigma |k\rangle\simeq -m_{k}$ is 
weakly coupled because the coordinate fields on $\cpN$ 
have large masses $|m_{i}-m_{k}|$.\footnote{The masses $m_{i}$ play the 
same r\^ole as Higgs vevs in four dimensional gauge theories, and the 
coordinate fields are like W bosons \cite{neu}.} The exact formula 
(\ref{vaceq}) then predicts that $\langle k|\sigma |k\rangle$ is 
given by an instanton series of the form
\begin{equation}
\label{insser}
\langle k|\sigma |k\rangle = -m_{k} + \sum_{j=1}^{\infty} c_{j}^{(k)} 
\Lambda^{j(N+1)}\, ,
\end{equation}
where the $j$-instanton contribution $c_{j}^{(k)}$ is a calculable 
function of the masses, for example $c_{1}^{(k)}=1/\prod_{j\not 
=k}(m_{j}-m_{k})$. Obviously, in the regime $|m_{i}-m_{k}|\gg 
|\Lambda|$, instanton calculus is valid:
the expectation value $\langle k|\sigma |k\rangle$ 
is a smooth, $2\pi$ periodic function of $\theta$,
\begin{equation}
\label{period}
\langle k|\sigma |k\rangle(\theta + 2\pi) = 
\langle k|\sigma |k\rangle(\theta)\, ,
\end{equation}
as are the other correlators in the $k$-th vacuum. 

When the masses $|m_{i}-m_{k}|$ decrease, the coupling grows, and we 
enter a regime where the semiclassical instanton calculus 
is no longer reliable. In our model, there is actually a sharp 
transition between an instanton dominated semiclassical regime and 
a purely strongly coupled regime. Mathematically, the transition 
occurs because series in $\Lambda^{N+1}$ like (\ref{insser}) have a finite
radius of convergence $R(m_{i})$. When $|\Lambda|^{N+1}>R(m_{i})$,
all the deductions based on a na\"\i ve discussion of instanton series 
can be invalidated.\footnote{For an 
analysis of the large $N$ limit of the analytic continuations, see 
\cite{fer1,fer2D}.} 
To be more specific, let us consider the case $N=1$, 
$m_{1}=-m_{2}=m$.
For $|m|>|\Lambda|$, the expectation values are given by convergent
instanton series deduced from (\ref{vaceq}),
\begin{equation}
\label{serex}
\langle 1|\sigma |1\rangle = -\langle 2|\sigma |2\rangle = -m 
\sum_{j=0}^{\infty} {(-1)^{j+1}\Gamma(j-1/2)\over 2\sqrt{\pi} j!} 
\left( {\Lambda\over m}\right)^{2j}\cdotp
\end{equation}
However, for $|m|<|\Lambda|$, one must use the analytic continuation to 
(\ref{serex}), given by
\begin{equation}
\label{n1case}
\langle 1|\sigma |1\rangle = -\langle 2|\sigma |2\rangle = 
-\sqrt{m^{2}+\Lambda^{2}}\, .
\end{equation}
A very important property of 
the analytic continuations is that they have branch cuts.
Equation (\ref{period}) is no longer valid and is replaced by
\begin{equation}
\label{periodfrac}
\langle\sigma\rangle_{1}(\theta + 2\pi) = 
\langle\sigma\rangle_{2}(\theta)\, .
\end{equation}
The statement for arbitrary $N$ is that the vacua are permuted for 
$\theta\rightarrow\theta +2\pi$ at strong coupling. Since for a 
generic choice of the masses $m_{i}$ the $N+1$ vacua are physically 
inequivalent, we see that in general {\it the physics is not 
$2\pi$ periodic in $\theta$.}\footnote{In the case $N=1$, 
the vacua $|1\rangle$ and $|2\rangle$ are physically equivalent due 
to a special ${\mathbb Z}_{2}$ symmetry of the theory, but for $N\geq 
2$ this does not occur.} Let us note that if we sum up over all the 
vacua, as would be the case in a calculation of the path integral in 
finite volume, then the $2\pi$ periodicity is restored. However, in 
infinite volume, cluster decomposition implies that
we should restrict ourselves to a given vacuum. In that case, though
the path integral is not $2\pi$ periodic, we still have 
\begin{equation}
\label{newperiod}
\langle\sigma\rangle_{k}\bigl((\theta + 2\pi (N+1)\bigr) = 
\langle\sigma\rangle_{k}\bigl( \theta\bigr)\, .
\end{equation}
This is consistent with a modified quantization law
\begin{equation}
\label{newqc}
(N+1) \int\! L_{\theta}/\theta \in {\mathbb Z}
\end{equation}
that is to be compared with the classical quantization law 
(\ref{qc}). Equation (\ref{newqc}) 
would correspond to so-called ``fractional instantons,'' 
field configurations with fractional topological charge $1/(N+1)$. 
The fractional instanton picture remains elusive, however, because we 
are not able to exhibit the corresponding field 
configurations.\footnote{They cannot be smooth configurations of the
microscopic fields, but might be naturally described in terms of 
composite fields that enter in the quantum effective action.} 

A precise way to state the result obtained above is to say that, at 
strong coupling, the 
transformation $\theta\rightarrow\theta +2\pi$ corresponds to a 
non-trivial monodromy of the vacua, and that, interestingly,
{\it this monodromy is not a symmetry transformation.} Another 
context where similar phenomena take place is four 
dimensional ${\cal N}=2$ gauge theories. The analogy with the $\cpN$ 
model discussed above was emphasized in \cite{Dorey}, and fractional 
instanton contributions were computed in the large $N$ limit in
\cite{fer1}. The formulas of \cite{fer1} clearly 
show that some observables transform in a complicated way under 
$\theta\rightarrow\theta +2\pi$. The 
transformation $\theta\rightarrow\theta +2\pi$ is actually implemented in 
${\cal N}=2$ gauge theories by a non-trivial duality transformation, 
which is a symmetry of the low energy effective action. Unlike the 
monodromy transformation of our two dimensional model, it could 
be a symmetry of the whole theory. To check this explicitly is 
however a highly non-trivial issue. 
\section*{Acknowledgements}
I would like to thank Adel Bilal and Edward Witten
for useful comments.
This work was supported in part by the Swiss National Science 
Foundation and by a Robert H.~Dicke fellowship.

\begin{thebibliography}{99}
%
%
\bibitem{adda}{A.~D'Adda, M.~L\"uscher and P.~Di Vecchia, 
\npb{146}{1978}{63}.}
%
\bibitem{coleman}{S.~Coleman, \ap{101}{1976}{239}.}
%
\bibitem{added}{E.~Witten, \ap{128}{1980}{363},\\
Y.~Oz and A.~Pasquinucci, \plb{444}{1998}{318},\\
E. Witten, \prl{81}{1998}{2862}.}
%
\bibitem{AG}{L.~\'Alvarez-Gaum\'e and D.Z.~Freedman, 
\cmp{80}{1981}{443},\\
L.~\'Alvarez-Gaum\'e and D.Z.~Freedman, \cmp{91}{1983}{87}.}
%
\bibitem{HH}{A.~Hanany and K.~Hori, \npb{513}{1998}{119}.}
%
\bibitem{Dorey}{N.~Dorey, \jhep{11}{1998}{005}.}
%
\bibitem{adda2}{A.~D'Adda, A.C.~Davis, P.~Di Vecchia and 
P.~Salomonson, \npb{222}{1983}{45}.}
%
\bibitem{neu}{F.~Ferrari, \plb{496}{2000}{212},\\
F.~Ferrari, \jhep{06}{2001}{057}.}
%
\bibitem{fer1}{F.~Ferrari, \npb{612}{2001}{151}.}
%
\bibitem{fer2D}{F.~Ferrari, {\it The large $N$ expansion, fractional 
instantons and double scaling limits in two dimensions,} NEIP-01-008,
PUPT-1997, LPTENS-01/11, in preparation.}
%
\end{thebibliography}
\end{document}